\documentclass[reprint, aps, prx, superscriptaddress]{revtex4-2}
\usepackage{graphicx}% Include figure files
\graphicspath{{E:/Dropbox (Personal)/BEC2/Arrow of time/Schematic figure/} {E:/Dropbox (Personal)/BEC2/Arrow of time/Full loop/Optimization graphs/} {E:/Dropbox (Personal)/BEC2/Arrow of time/Full Loop SGI paper/Figures/} {E:/Dropbox (Personal)/BEC2/Arrow of time/arXiv/} {C:/Users/BEC2M/Dropbox (Personal)/BEC2/Arrow of time/Schematic figure/} {C:/Users/BEC2M/Dropbox (Personal)/BEC2/Arrow of time/Full loop/Optimization graphs/} {C:/Users/BEC2M/Dropbox (Personal)/BEC2/Arrow of time/Full Loop SGI paper/Figures/} {C:/Users/BEC2M/Dropbox (Personal)/BEC2/Arrow of time/arXiv/} {C:/Users/Yair Margalit/Dropbox (Personal)/BEC2/Arrow of time/Schematic figure/} {C:/Users/Yair Margalit/Dropbox (Personal)/BEC2/Arrow of time/Full loop/Optimization graphs/} {C:/Users/Yair Margalit/Dropbox (Personal)/BEC2/Arrow of time/Full Loop SGI paper/Figures/} {C:/Users/Yair Margalit/Dropbox (Personal)/BEC2/Arrow of time/arXiv/}}
\usepackage{dcolumn}% Align table columns on decimal point
\usepackage{bm}% bold math
\usepackage{setspace}
\usepackage{booktabs}
\usepackage{amsmath}
\usepackage{ulem} % CH: strike out \sout{some text}

\usepackage{xcolor}
\definecolor{darkGreen}{rgb}{0, 0.7, 0}

\newcommand{\be}{\begin{equation}}
\newcommand{\ee}{\end{equation}}

\newcommand{\ket}{\rangle}

\begin{document}

\title{Realization of a complete Stern-Gerlach interferometer:\\
 Towards a test of quantum gravity}

\author{Yair Margalit}
\thanks{Corresponding author: margalya@bgu.ac.il; Current address: Research Laboratory of Electronics, MIT-Harvard Center for Ultracold Atoms, Department of Physics, Massachusetts Institute of Technology, Cambridge, Massachusetts 02139, USA.}
\author{Or Dobkowski}
\author{Zhifan Zhou}
\author{Omer Amit}
\author{Yonathan Japha}
\author{Samuel Moukouri}
\author{Daniel Rohrlich}
	\address{Department of Physics, Ben-Gurion University of the Negev, Be'er Sheva 84105, Israel}
\author{Anupam Mazumdar}
    \address{Van Swinderen Institute, University of Groningen, 9747 AG Groningen, The Netherlands}
\author{Sougato Bose}
	\address{Department of Physics and Astronomy, University College London, Gower Street, WC1E 6BT London, United Kingdom}
\author{Carsten Henkel}
	\address{Institute of Physics and Astronomy, University of Potsdam, Germany}
\author{Ron Folman}
	\address{Department of Physics, Ben-Gurion University of the Negev, Be'er Sheva 84105, Israel}
%	\email{margalya@post.bgu.ac.il}

\begin{abstract}
The Stern-Gerlach effect, discovered a century ago, has become a paradigm of quantum mechanics. Surprisingly there has been little evidence that the original scheme with freely propagating atoms exposed to gradients from macroscopic magnets is a fully coherent quantum process. Specifically, no full-loop Stern-Gerlach interferometer has been realized with the scheme as envisioned decades ago. Furthermore, several theoretical studies have explained why such an interferometer is a formidable challenge. Here we provide a detailed account of the first full-loop Stern-Gerlach interferometer realization, based on highly accurate magnetic fields, originating from an atom chip, that ensure coherent operation within strict constraints described by previous theoretical analyses. Achieving this high level of control over magnetic gradients is expected to facilitate technological as well as fundamental applications, such as probing the interface of quantum mechanics and gravity. While the experimental realization described here is for a single atom, future challenges would benefit from utilizing macroscopic objects doped with a single spin. Specifically, we show that such an experiment is in principle feasible, opening the door to a new era of fundamental probes.
\end{abstract}

\maketitle

\section*{Introduction}
The discovery of the Stern-Gerlach (SG) effect\,\cite{stern-gerlach,SLB} was followed by ideas concerning a full-loop SG interferometer (SGI) consisting of a beam of atoms exposed to field gradients from macroscopic magnets \,\cite{Wigner}. However, starting with Heisenberg, Bohm and Wigner \cite{briegel} a coherent SGI was considered
impractical because it was thought that the macroscopic device could not be accurate enough to ensure a reversible splitting process, namely, a complete overlap in position and momentum of the two interferometric paths.
Bohm, for example, noted that the magnet would need to have ``fantastic" accuracy\,\cite{Bohm}.
Englert, Schwinger and Scully analyzed the problem in more detail and coined it the Humpty-Dumpty (HD)
effect \cite{ESS_1,ESS_2,ESS_3, englert_1}. They too concluded that for
significant coherence to be observed, exceptional accuracy in controlling magnetic fields would be required. Indeed, while atom interferometers based on light beam-splitters enjoy the quantum accuracy of the photon momentum transfer\,\cite{PritchardReview}, the SGI magnets not only have no such quantum discreteness, they also suffer from inherent lack of flatness due to Maxwell's equations.
Later work added the effect of dissipation and suggested that only low-temperature magnetic field sources would enable an operational SGI\,\cite{caldeira}. Claims have even been made that no coherent splitting is possible at all \cite{Devereux}.

\begin{figure}
\centerline{
\includegraphics*[angle=0, width=\columnwidth]{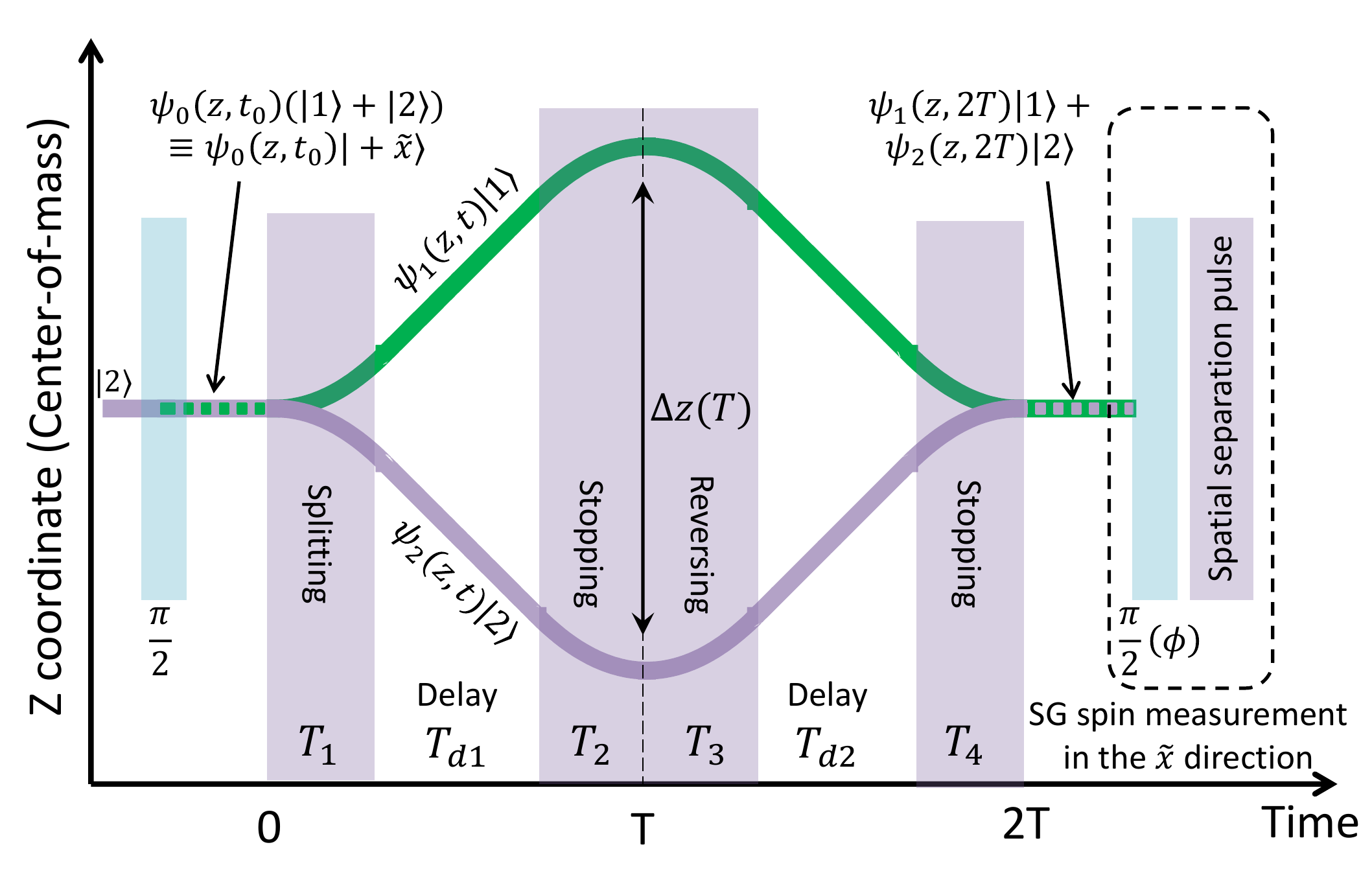}}
\caption{The longitudinal full-loop SGI ($z$ position vs. time). The figure is plotted in the center-of-mass frame. The interferometer operates for a duration of 2T (on the order of a few hundred $\mu s$), and consists of four magnetic gradient pulses (purple columns). The states $|1\ket\equiv|F=2,m_F=1\ket$ and $|2\ket\equiv|2,2\ket$ are defined along the $\tilde{z}$ axis in the Bloch sphere (different from real-space coordinates). The signal is made of spin population fringes. The experiment starts with a spin in the $\tilde{x}$ direction, and the final measurement is again of the spin in the $\tilde{x}$ direction. The latter is performed by mapping the spins from $\tilde{x}$ to $\tilde{z}$ with a $\pi/2$ rotation and applying a SG gradient to separate the populations before taking an image.
The same configuration may be used for a macroscopic-object interferometer, see Section: Testing Different Aspects of Gravity.
\label{fig:schematic}}
\end{figure}

Here we detail the realization of a full-loop SGI, in which magnetic field gradients act on the atom during its flight through the interferometer, as originally envisioned. We obtain a high full-loop SGI contrast of 95\% with a spin interference signal. We achieve this utilizing the highly accurate magnetic fields of an atom chip\,\cite{keil}. Following the footsteps of impressive endeavors\,\cite{OldSG0,OldSG2,OldSG3,OldSG4,OldSG5,OldSG6, OldSG7,OldSG8,OldSG9,OldSG10, Marechal2000}, and additional recent scientific advancements with spatial SGIs \cite{machluf, spatialSG}, this is, to the best of our knowledge, the first realization of a complete SGI interferometer analogous to that originally envisioned.

Achieving this high level of control over magnetic gradients may facilitate fundamental research as well as technological applications. SG interferometry with mesoscopic objects has been suggested as a compact detector for space-time metric and curvature \cite{MIMAC}, possibly enabling detection of gravitational waves. It has also been suggested as a probe for the quantum nature of gravity\,\cite{Bose2017}. Such SG capabilities may also enable searches for exotic effects like the fifth force, or the hypothesized self-gravitation interaction\,\cite{self-gravity, self-gravity2, self-gravity3}. In the following we show that such an experiment is in principle feasible.

We note that the realization presented here has already enabled the construction of a unique matter-wave interferometer whose phase scales with the cube of the time the atom spends in the interferometer \cite{T3SGI}. This realization has been suggested as an experimental test for the Einstein's equivalence principle when extended to the quantum domain \cite{applications3}.

High magnetic stability and accuracy may also make possible technological and metrological applications such as large-area atom interferometry\,\cite{LMT}, sensitive probing of electron transport, e.g., squeezed currents\,\cite{SubShotNoise}, as well as nuclear magnetic resonance and compact accelerators\,\cite{applications2}. We note that as the SGI makes no use of light, it may serve as a high-precision surface probe at short distances for which administering light is hard. In addition, our atom chip setup is compatible with cryogenic environments, and may hence probe cold surfaces, while laser light may cause unwanted heating.

Finally, let us emphasize that a full-loop scheme, with its spin-population observable, as described here, has several advantages for macroscopic-object interferometry, over other suggestions which utilize a spatial interference pattern as their signal. First, a spin-population observable requires no high-resolution imaging. For massive objects this may be crucial. Second, as shown in\,\cite{T3SGI}, a SGI can achieve a $T^3$ scaling of the phase accumulation, enabling high sensitivity. Furthermore, the SGI allows us to apply magnetic forces throughout the interferometer, and a significant splitting may be achieved in a few ms or so. Last, and perhaps most importantly, the spin-population observable does not require free propagation to develop, in contrast to a spatial interference signal. These reasons provide a crucial advantage when taking into account the high decoherence rate expected for macroscopic objects.

\section*{The full-loop SG interferometer}
A schematic of the full-loop interferometer is presented in Fig.\,\ref{fig:schematic}. The full-loop SGI utilizes four magnetic gradient regions or pulses: splitting, stopping, accelerating back and stopping, the latter two closing a loop in the space-time diagram. To the best of our knowledge, all previous SG realizations used only the first one or two operations, thereby realizing at most a `half loop'. The entanglement of spin with the spatial degrees of freedom persists throughout the SGI, and the magnetic moment associated with the spin is used to control the external degrees of freedom, using magnetic gradients and the SG effect. The SGI actively recombines the wavepackets in both position and momentum, and uses the spin state of the recombined wavepacket as an interference signal. This is in contrast to the spatial fringe SG interferometer realized in our previous work\,\cite{machluf, spatialSG} which consists only of splitting and stopping the wavepackets (thus corresponding to a half loop, to which we added spin mixing to allow for spatial fringes to form, in analogy to a double-slit experiment).

The recombination in the full-loop SGI, is in fact required to be a time-reversal operation of the splitting process, such that the final two magnetic gradients exactly undo the first two. In order to obtain high coherence (or contrast) in the output of a spatial interferometer, one must apply stable and accurate operations on the atom, such that the final relative distance between the wavepackets' centers, $\Delta z(2T)$, and the final relative momentum between the centers, $\Delta p(2T)$, are minimized, where $2T$ is the interferometer duration. Inaccuracy of the magnetic field gradient throughout the particle's trajectory gives rise to imperfect overlap, either in position or momentum, and will cause a decay in the resulting interferometric contrast.

The difficulty in maintaining spin coherence due to inaccuracy is illustrated by the following simple argument relating to the momentum splitting (as argued by Heisenberg and others\,\cite{ESS_1,briegel, Heisenberg}). In order to achieve macroscopic splitting of the wavepackets using a differential force $F$ acting for a duration $T_1$, the relative momentum between the split wavepackets $FT_1$ must be much bigger than the initial wavepacket width in momentum $\sigma_p$, i.e. $FT_1 \gg \sigma_p$. As each part of the wavepacket samples a different part of the potential, it acquires a different phase. The linear phase spread over the wavepacket is thus given by
$|\delta\phi| = |\delta (ET_1/\hbar)| = |(\partial E/\partial z) \sigma_z T_1/\hbar| = FT_1 \sigma_z/\hbar \gg \sigma_p \sigma_z/\hbar$, where $\sigma_z$ is the initial wavepacket width in position ($\sigma_p$ and $\sigma_z$ are both defined in the $z$ direction), and $E$ is the atom's energy due to the magnetic field. 
By invoking the uncertainty principle, one finds that $\delta\phi$ may not be made small, further complicating the recombination, where a successful recombination requires maximizing the overlap integral of the two wavepackets. In other words, large splitting requires large relative momentum in units of the internal momentum width, which corresponds to a large phase spread over the wavepacket (due to the relation between momentum and phase - $e^{ipz/\hbar}$).
To achieve high coherence, this phase spread originating from momentum splitting has to be undone. If the size of the two wavepackets is the same, minimizing $\Delta p$ ensures to some degree that the phase profile of both wavepackets is the same, which is sufficient. Note that in a practical experiment, the phase pattern is actually more complex and harder to match due to the curvature of the magnetic potential.

The precision with which one has to recombine the wavepackets is set by the spatial coherence length $l_z$ and momentum coherence width $l_p$, which we use as a phenomenological model to describe the loss of contrast. These measures of coherence are inversely proportional to the momentum and position uncertainties of the atom, $\sigma_p$ and $\sigma_z$, and may be defined as \cite{JorgArxiv}
\begin{equation}\label{eq:coherence length}
l_z=\frac{\hbar}{\sigma_p}, \quad l_p=\frac{\hbar}{\sigma_z}, \end{equation}
where the momentum and position uncertainties satisfy the uncertainty relation $\sigma_p\sigma_z\geq \hbar/2$.
If the two paths at the output port of the interferometer (time $t=2T$) are displaced by a distance $\Delta z(2T)$ the contrast is expected to drop as $C\propto \exp[-\frac12 \left(\Delta z(2T)/l_z \right)^2]$. Equivalently, if the two paths are displaced in momentum by $\Delta p(2T)$ the contrast reduces as $C\propto \exp[-\frac12 \left(\Delta p(2T)/l_p \right)^2]$.

In the case of a minimal-uncertainty Gaussian wavepacket with a negligible expansion rate, the loss of contrast (or coherence) $C$ due to inaccuracy of the recombination process is quantified by the HD equation, which is given by\,\cite{ESS_2}
\begin{equation}\label{eq:HD}
C = \exp\left[ -\frac{1}{8}\left(\frac{\Delta z(2T)}{\sigma_z}\right)^2 - \frac{1}{8}\left(\frac{\Delta p(2T)}{\sigma_p}\right)^2 \right].
\end{equation}
(Note a factor $1/2$ in our definition of $\Delta z$ and $\Delta p$ relative to the original definition).
This equation is the result of calculating the overlap integral between the Gaussian wavefunctions at the end of the interferometer.
In order to keep the contrast close to unity, a `microscopic' level of accuracy is required at the end of the interferometer,
described by the relations $|\Delta z(2T)|\ll\sigma_{z}$ and $|\Delta p(2T)|\ll\sigma_p$.
Quantitatively, to maintain a contrast of $ \simeq 0.99$ using $FT_1 / \sigma_p = 10^3$, it turns out that one must control the fields with an accuracy of $\delta F/F = 10^{-5}$ \cite{ESS_1, ESS_2}, a formidable technical challenge. Addressing this challenge is expected to open the door to a wide variety of new experiments for technology and fundamental studies.

\section*{Experiment}

Our experiment begins by releasing a BEC of about $10^4$ $^{87}$Rb atoms from a magnetic trap below an atom chip\,\cite{keil}. We initially prepare the BEC in the state $\vert F, m_F\rangle =\vert 2,2\rangle$, and then create a superposition of the two spin states $|F,m_F\rangle=|2,2\rangle \equiv |2\rangle$ and $|2,1\rangle \equiv |1\rangle$ by applying a $\pi/2$ RF pulse. These two states constitute an effective two-level system, as all other states in the $F=2$ manifold are pushed out of resonance by the nonlinear Zeeman shift generated using an external bias field (see Methods for more details). To avoid dephasing of the spin superposition due to noise in the bias fields, we add $\pi$ pulses giving rise to an echo sequence (see Methods). The full-loop SGI is then applied by using a series of four magnetic gradient pulses (gradients along the axis of gravity, $z$, see Fig.\,\ref{fig:schematic}), which are generated by running currents on the atom chip (more details on the setup can be found in Ref.\,\cite{spatialSG}). The first pulse, of duration $T_1$, splits the superposition into two momentum components, which then freely propagate during a delay time $T_{d1}$. The wavepackets are then stopped relative to one another (pulse duration $T_2$), accelerated back (pulse duration $T_3$), and after a second delay time $T_{d2}$ are stopped again (pulse duration $T_4$), ideally when overlap in space and momentum is maximal. As the direction of acceleration in the second and third pulses is opposite to that of the first and fourth pulses, we name the $T_2$ and $T_3$ pulses, reverse pulses. As the $T_3$ and $T_4$ pulses are required to undo the splitting in position and momentum created by the $T_1$ and $T_2$ pulses, $T_3$ and $T_4$ are named the recombination pulses.
We obtain the population signal with the help of a second $\pi/2$ pulse followed by a spin population measurement. We measure the visibility by scanning the phase $\phi$ of the second $\pi/2$ pulse (Fig.\,1), and observe the contrast of the resulting population fringes.

\begin{figure}
\centerline{
\includegraphics*[angle=0, width=\columnwidth]{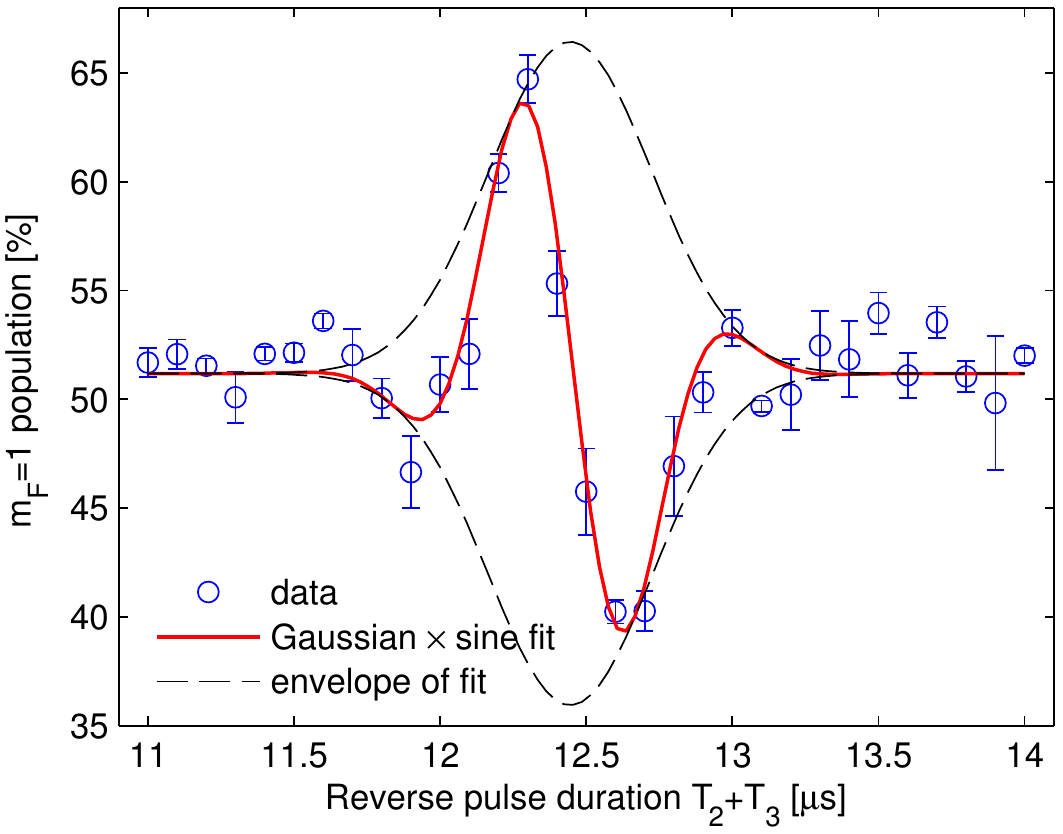}}
\caption{Full-loop optimization procedure: population output as a function of the reverse pulse duration $T_2+T_3$, for $T_1 = T_4 = 6\,\mu$s, $T_{d1}=300\,\mu$s, and a relative acceleration $a=635$\,m/s$^{2}$. The population oscillates around the optimal point as expected by a simplified model of a Gaussian times a sine. The peak of the Gaussian envelope corresponds to the time at which the wavepackets' overlap integral at the end of the interferometer is maximized for the given parameters. The sine function corresponds to the added phase between the two interferometer arms, per unit time of reversing pulse.
\label{fig:optimization}}
\end{figure}

In order to minimize $\Delta z(2T)$ and $\Delta p(2T)$ (inaccuracies in the final recombination), and thus maximize the visibility of the interference signal, we must optimize the experimental parameters. In Fig.\,\ref{fig:optimization} we present an example optimization run, in which we fix the durations of the first and last gradient pulses $T_1$ and $T_4$ and also the durations of the delay times $T_{d1}$ and $T_{d2}$ (usually $T_1 = T_4$ and $T_{d1} = T_{d2}=T_d$ to begin with). We measure the output population as a function of the duration $T_2+T_3$ of the second and third gradient (reverse) pulses, while keeping the total duration $T_2+T_3+T_{d2}$ constant. The data nicely fit a Gaussian envelope times a sine function. The peak of the envelope corresponds to the point in which the overlap integral is maximal. Ideally for linear magnetic gradients, one would expect the peak overlap to occur when the sequence is symmetric i.e. $T_1 + T_4 = T_2+T_3$ (assuming $T_{d1} = T_{d2}$). However the non-linearity of the magnetic potential created by the chip wires in the $z$ direction\,\cite{spatialSG}, together with the wavepackets' movement with respect to the chip, breaks the symmetric timing diagram. The optimal duration of the gradient pulses depends on the delay times, the initial distance from the chip, and the specific scheme used (see Methods section for more details about the optimization procedure).
Generally, one needs to optimize two independent parameters - one to minimize the final relative position $\Delta z(2T)$, and one to minimize the final relative momentum $\Delta p(2T)$.

While, as noted in the introduction, accuracy is the main challenge, we also need to address the issue of stability, whereby temporal fluctuations may give rise to dephasing, decoherence and drifts.
%Instability in the case of the full-loop SGI originates from temporal fluctuations and technical drifts due to the environment (either the magnet itself or beyond it).
Even in the absence of decoherence, noise may cause the interference phase to jitter from one experimental shot to the next (e.g. due to a fluctuating bias field), thus dephasing the averaged phase, or preventing recombination altogether in the case of a different noise (e.g. fluctuating gradients).
In this work we achieve high stability by utilizing an atom chip\,\cite{keil} with several advantages, including strong magnetic gradients so that the experimental duration is very short and consequently the interaction with external noise is brief, and very low inductance so that the gradients can be switched in micro-seconds.
In addition, the structure and position of the magnet are very precise as it is made of a near-perfect wire.
%Furthermore, relative to our previous work where low visibility was obtained in a spatial SG configuration\,\cite{machluf},
Also, care was taken to reduce a wide range of hindering effects. For example, a novel method is used to reduce the effect of current fluctuations on the chip by utilizing a 3-wire configuration which produces a quadrupole and exposes the wavepackets to a weaker magnetic field from the chip,
while maintaining strong gradients. Using these advantages, we have been able to show low phase noise\,\cite{spatialSG} (standard deviation as low as 0.1\,rad of the spatial fringe SGI), demonstrating the stability of the apparatus.

\section*{Results}
%%% Proving recombination

\begin{figure}
\centerline{
\includegraphics[width=\columnwidth]{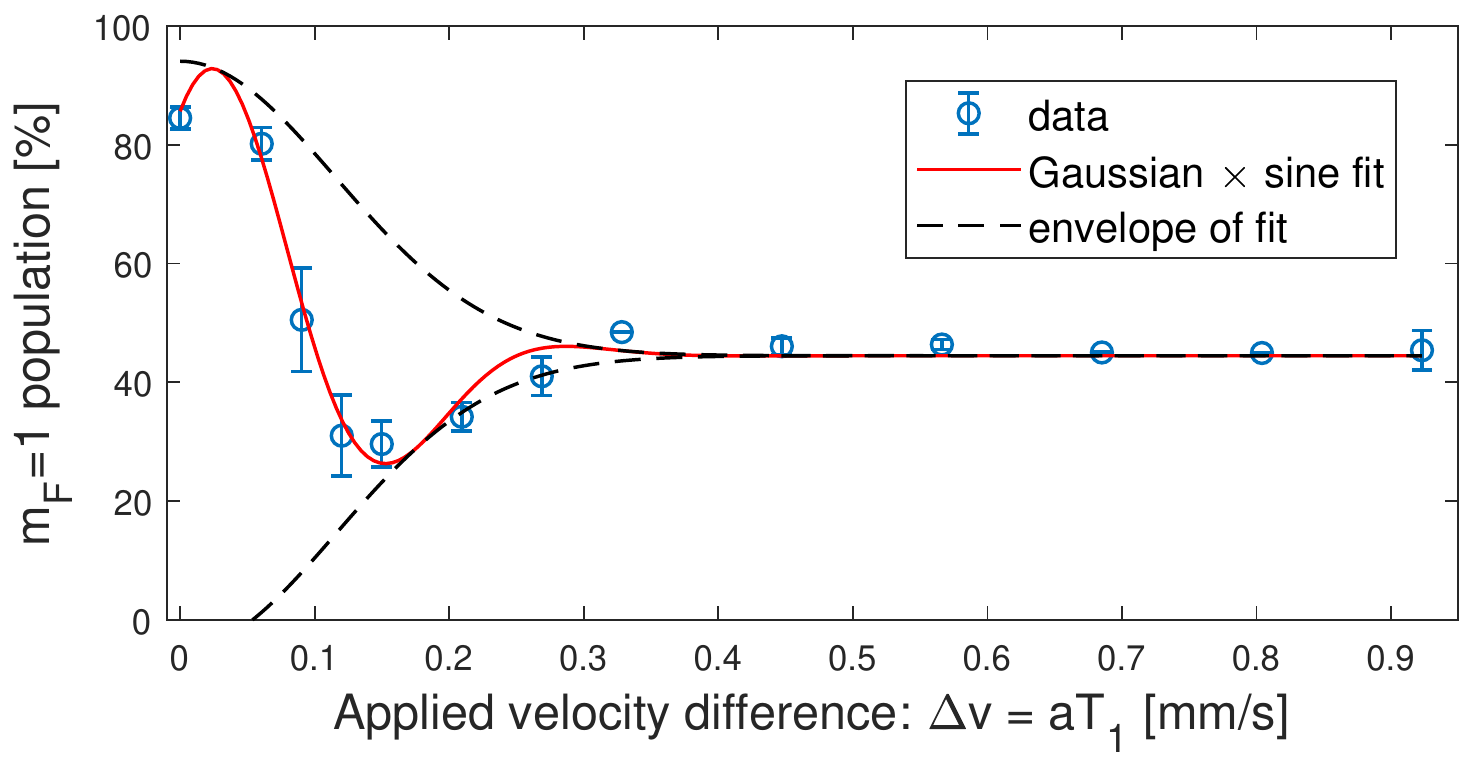}}
\caption{\label{fig:single kick} Single kick effect on the contrast: population as a function of the applied velocity difference $\Delta v$ between the wavepakcets for a single pulse of duration $T_1$ (where we set $T_2 = T_3 = T_4=0$). As the splitting is increased, the population decays to 50\% - corresponding to zero contrast. In this measurement we take care that the spatial separation is as small as possible ($\Delta z \leq 50 $ nm) such that the decay due to spatial splitting is negligible. The velocity difference is calculated according to $\Delta v = a T_1$, where $a=59.5$\,m/s$^{2}$ is the applied relative acceleration. The data are fitted to a Gaussian times a sine function, and the fit returns a momentum coherence width of $l_p/m = 0.12 \pm 0.03$\,mm/s, where $m$ is the atom's mass. To the best of our knowledge, this is the first direct measurement of the momentum coherence width of a BEC (see previous results using neutrons\,\cite{NeutronMomentumCoherence} and atomic beams\,\cite{BeamMomentumCoherence}).}
\end{figure}

\begin{figure}
\centerline{
\includegraphics[width=\columnwidth]{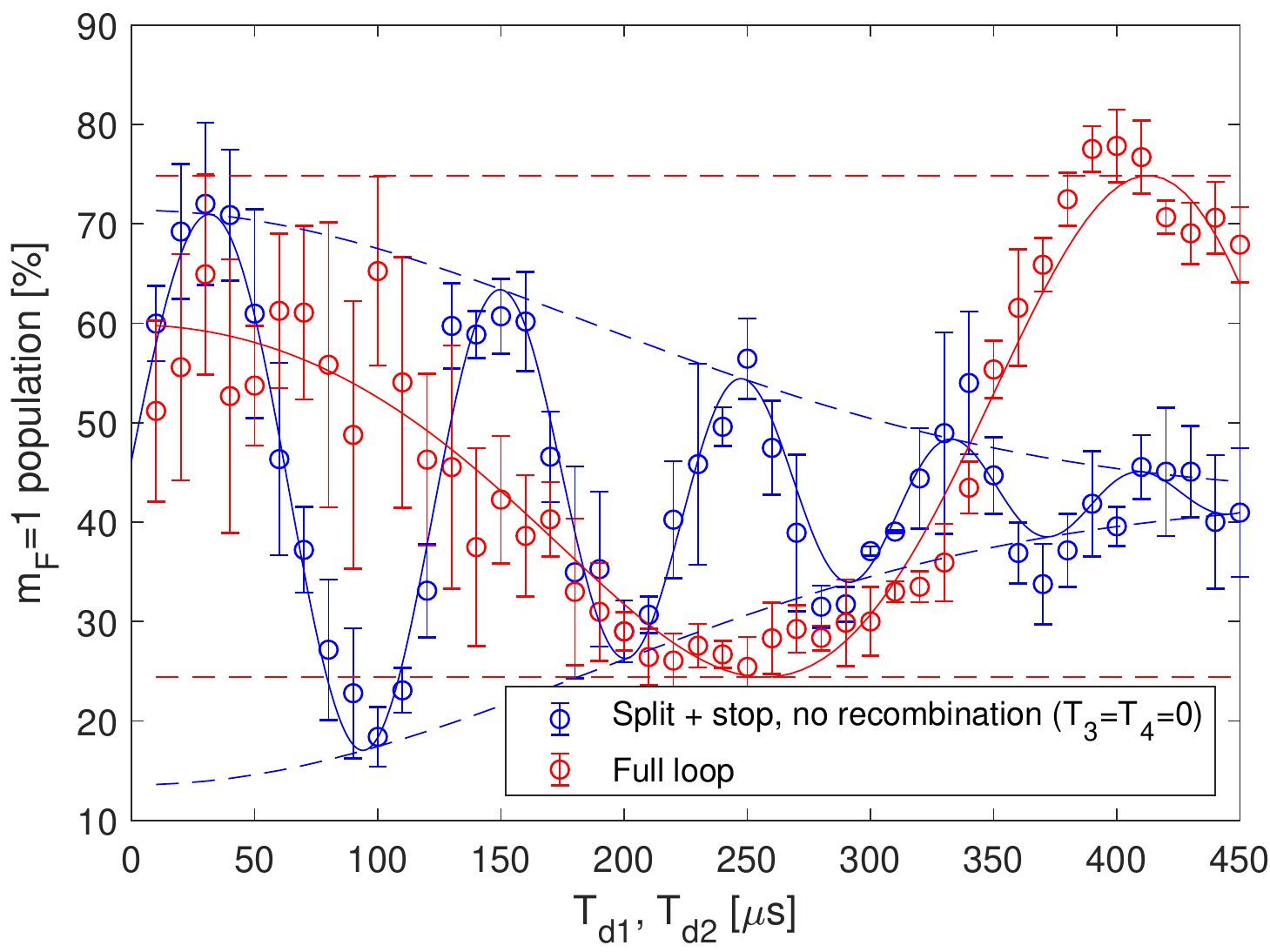}}
\caption{\label{fig:recombination} Full-loop recombination: population in the $m_F=1$ state as a function of delay times $T_d$ for two different sequences, using $T_1 = 5.4\,\mu$s and a relative acceleration $a=481$\,m/s$^{2}$. Blue data is an optimized splitting ($T_1$) and stopping ($T_2$) sequence without recombination (i.e. $T_3,\, T_4 = 0$). The optimization procedure ensures that the visibility loss due to momentum splitting is minimized for every value of $T_{d1}$, meaning we see visibility loss mainly due to spatial splitting (as it becomes large relative to the coherence length). The red data use the same parameters of $T_1,\, T_{d1}$, and $T_2$, but add the recombining pulses. Here we use $T_{d1} = T_{d2}=T_d$, and they are scanned together. Solid lines are fits to the data (see text for details); dashed lines represent the envelope of the fit, demonstrating the visibility as a function of the delay time. Note that the red data seems to start with a lower contrast. This effect of flatness and ``increasing contrast" is expected by the theory\,\cite{T3SGI}, and originates from a non-linear phase term. One can clearly see that for long delay times, the spatial splitting reduces the the visibility of the interference. The recombination then revives the visibility, demonstrating successful recombination in both position and momentum.}
\end{figure}

We now validate that we are able to successfully recombine in position and momentum, as both can cause loss of visibility.
First, in Fig.\,\ref{fig:single kick}, we present the loss of coherence due to the first magnetic pulse alone (i.e. by setting $T_2,\,T_3,\,T_4=0$), giving rise to orthogonality in momentum.
The contrast drops to $1/\sqrt{e}$ at $l_p/m=0.12$\,mm/s momentum splitting. Next, we apply a much stronger momentum splitting of $\Delta p/m = 2.6$\,mm/s, and show in Fig.\,\ref{fig:recombination} (blue data) that we are able to undo this orthogonality by a second gradient pulse, which stops the relative motion of the wavepackets. However, as we extend the delay time $T_{d1}$ the wavepackets start to split in space and we observe a decay in visibility which cannot be restored using $T_2$ alone, as the distance between the wavepackets becomes larger than their coherence length $l_z$. We validate that this loss of visibility is mainly due to spatial splitting, by
optimizing $T_2$ for each value of $T_{d1}$, such that maximal visibility is achieved (see Methods for details).

Using this method we are able to also accurately determine $l_z$.
To describe the expected loss of contrast with increasing spatial splitting, we fit the blue data with a Gaussian times a sine function of the form $P_1 = A\exp(-\frac12 T_d^2/\tau^2)\sin[\delta\phi(T_d) + \phi_0]+c$, where $\delta\phi(T_d)$ is the accumulated phase difference, which contains terms proportional to $T_d$ and $T_d^2$ \cite{T3SGI}, $\phi_0$ is a constant phase shift, $A$ is the amplitude, $c$ is a constant, and $\tau$, the decay constant, is essentially the coherence time (as $T_d$ is much larger than the pulse time). Using the value of $\tau$ we calculate the spatial coherence length $l_z$ and obtain $l_z = 0.5 \pm 0.07\,\mu$m (see Methods). Note that the blue data decay to less than the expected 50\% value, probably due to imperfect RF $\pi/2$ pulses, affecting the state preparation and population measurement.

We are now ready to recombine the two wave packets. We add the recombining pulses $T_3$ and $T_4$ to the previous sequence, and generate the red data in Fig.\,\ref{fig:recombination}, where the value of $T_4$ is also optimized in a similar manner for maximal visibility. One can clearly see the Gaussian decay of the visibility due to spatial splitting in the first (blue) data, and the revival of the visibility due to the successful recombination of the spatially-split wavepackets (red data). At $T_{d1} = T_{d2}=350\,\mu$s, for example, the blue data decay to 16\% of their original amplitude, while the red data show no decay at this time scale.
The red data are fitted with a similar function as that used for the blue data, but without the decaying Gaussian term (as the decay is not visible in this range), namely $P_1 = A\sin[\delta\phi(T_d) + \phi_0]+c$.

In summary, we have clearly shown the successful recombination in momentum and position, thus realizing a complete SGI. We have also measured $l_p$ and $l_z$, and the observed visibility is proof that the accuracy of our recombination in momentum and position is better than these coherence scales (see Methods for more discussion on the results).

%\begin{tabular}{c}
% \includegraphics[width=\columnwidth]{{"E:/Dropbox (Personal)/BEC2/Arrow of time/PRL resubmission/Full_loop_contrast_vs_Dz_over_sigmaz_3_data_sets_v4"}.pdf}\\
% \includegraphics[width=\columnwidth]{{"E:/Dropbox (Personal)/BEC2/Arrow of time/PRL resubmission/Full_loop_contrast_vs_Dv_over_sigmav_3_data_sets_v4"}.pdf}
%\end{tabular}}

%left bottom right top
% \includegraphics[width=\columnwidth, trim=32 1 18 8, clip]{Full_loop_contrast_vs_Dz_over_sigmaz_3_data_sets_v7.pdf}\\
% \includegraphics[width=\columnwidth, trim=32 1 18 8, clip]{Full_loop_contrast_vs_Dv_over_sigmav_3_data_sets_v7.pdf}
\begin{figure}
\centerline{
\begin{tabular}{c}
 \includegraphics[width=0.9\columnwidth]{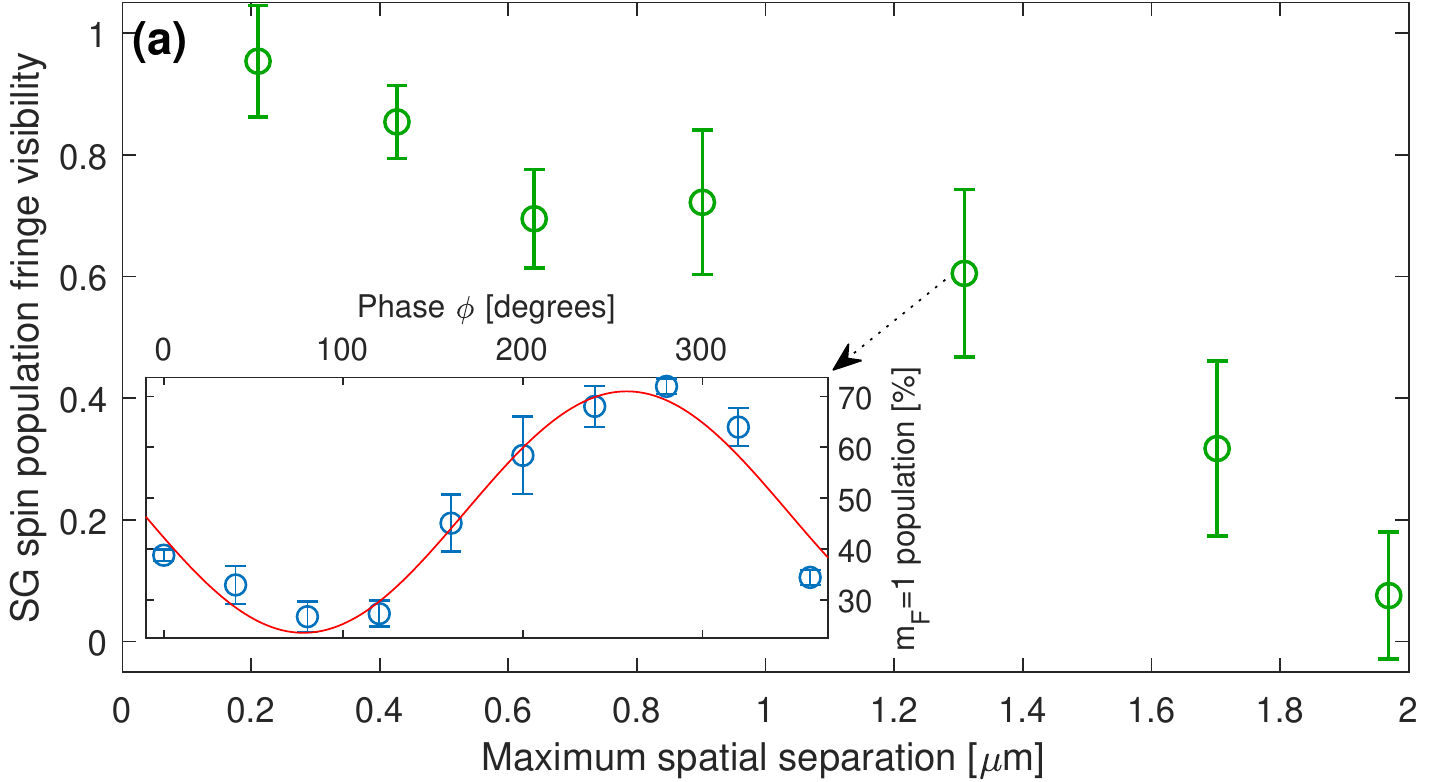}\\
 \includegraphics[width=0.9\columnwidth]{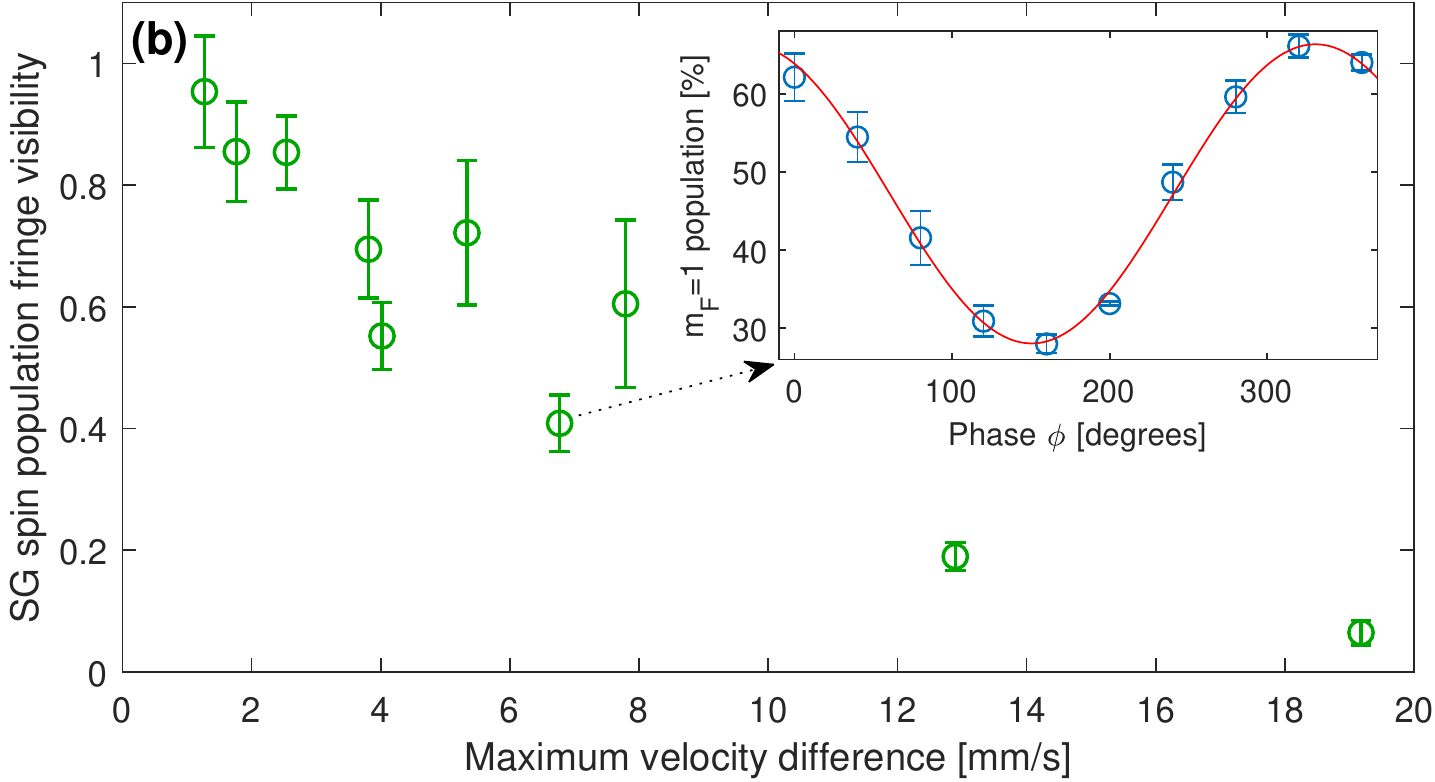}
\end{tabular}}
\caption{Analysis of accuracy:
(a) Spin population visibility vs. maximal separation $\Delta z(T)$. Experimental parameters are $T_1=2-30\,\mu$s, $T_{d1}\simeq T_{d2}=50-160\,\mu$s.
(b) Visibility vs. maximal momentum splitting $\Delta p(T_1)$. Experimental parameters are $T_1=2-6\,\mu$s, $T_{d1} \simeq T_{d2}=164-367\,\mu$s. In both (a) and (b), the relative acceleration is varied in the range $a=635-2641\,$m/s$^{2}$.
As expected, coherence decays as we increase spatial and momentum splitting. A detailed quantitative analysis could not explain the observed slope (see discussion in text).
In both panels, the visibility is normalized to the Ramsey visibility without splitting (i.e., no magnetic gradients), typically $\sim90\%$.
Insets show the raw data for two of the data points, namely the population fringe generated using a phase scan of the second $\pi/2$ pulse, and a fit to a sine wave used to extract the contrast (red line, see Methods). Data points in (a) and (b) are of different data sets.
\label{fig:contrast}}
\end{figure}

\section*{Limits on Accuracy}

To test the limits of our accuracy, in Fig.\,\ref{fig:contrast} we plot the visibility as a function of the maximum splitting in momentum $\Delta p(T_1)$ and in position $\Delta z(T)$.
In order to try and maximize the splitting, we utilize several different configurations: we invert the sign of the relative acceleration by reversing the sign of the currents in the chip wires, while in other sequences we keep the same currents while inverting the spins with the help of $\pi$ pulses (see Methods).
We also utilize a variety of magnetic gradient magnitudes by varying the current on the chip, and scan both the splitting gradient pulse duration $T_1$, and the delay time between the pulses $T_d$.
Each point in Fig.\,\ref{fig:contrast} was taken using different parameters, and was optimized independently. For weak splitting we observe high visibility ($\sim95\%$), while for a momentum splitting equivalent to 1\,$\hbar k$ (optical photon recoil on the rubidium D2 line, $\hbar k/m=5.8\,$mm/s, where $m$ is the atom's mass) the visibility is still high ($\sim75\%$) indicating that the magnetic field accuracy allowed reversing the splitting to a high degree. The quoted values are normalized to that of a pure Ramsey sequence, i.e. a sequence without any magnetic gradients, in order to cancel technical effects (see Methods).
The visibility as a function of maximum splitting is qualitatively the same for space and momentum splitting, and shows a decrease in visibility as splitting is increased.

%%%%%%%%

As we know the potential created by the chip, the wavepackets' positions and widths can be calculated in every step of the experiment. The HD equation (Eq.\,\ref{eq:HD}) should thus, in principle, allow for a quantitative calculation of the contrast using the experimental parameters. However, the equation makes two simplifying assumptions: first, that the wavepackets do not expand during the propagation through the interferometer, which does not hold in our case. Second, it assumes that there is no non-linear phase imprinted on the wavepackets (i.e. magnetic gradients are linear). In our case, phase non-linearity originates from both the curvature of the magnetic gradient generated by the chip, and from conversion of the mean-field potential energy of the BEC released from trap to kinetic energy (causing a quadratically-varying phase to evolve\,\cite{Phillips}). In order to account for these effects and try to quantitatively explain the loss of contrast in the full-loop SGI, we have developed a generalized wavepacket model for studying coherence of matter-wave interferometers\,\cite{Japha2019}. While we have simulated our experimental conditions with care, and while such simulations accurately describe our previous interferometry results (e.g. \cite{machluf, margalit, T3SGI, zhou, spatialSG}), the coherence drop observed in the full-loop experiment is not well described by the generalized wavepacket model\,\cite{Japha2019}, and neither by the HD model. Using the experimental parameters to calculate the visibility, both models predict values which are significantly higher than those observed in the experiment. %Noting that our current accuracy is 0.1\%, and our timing accuracy is similar, we expect the drop to be much weaker.
We therefore leave the quantitative comparison to future work.

In order to give limits on our accuracy in recombining the wavepackets, we use the experimentally measured coherence scales as a figure of merit, and examine the ratio between the achieved maximum splittings and the coherence scales. We are able to recombine wavepackets with a maximum spatial splitting of ${\Delta z(T)}/l_z=3.9$ before visibility goes to zero, and a maximum momentum splitting of ${\Delta p(T_1)}/l_p=158$. These results indicate that we are more successful in recombining in momentum than in space. The reason for this difference is yet unclear.

\section*{Testing different aspects of gravity}

The idea of using the SG interferometer with a macroscopic object as a probe for gravity has been described in detail in several theoretical works \cite{MIMAC,Bose2017,BoseNanoObject,Anupam2019,Marletto2017,applications3}.
These works detail a wide range of experiments from detection of gravitational waves to the testing of the quantum nature of gravity. Here, we discuss the feasibility of such an experiment using the full-loop SGI, as described above for an atom, and show that such an experiment using a macroscopic body is also feasible. As before, our interference observable is oscillations of spin population, rather than spatial fringes (density modulations). This observable, as demonstrated in the atomic SGI described above, has the advantages that there is no requirement for long evolution times in order to allow the spatial fringes to develop, there is no need for high resolution imaging to resolve the spatial fringes, and the phase accumulation may be fast (e.g. $T^3$ \cite{T3SGI}). Let us note that there are other proposals to realize a spatial superposition of macroscopic objects\,\cite{Romero5,Romero6}, but they do not rely on the SG effect, and they utilize spatial rather than spin population fringes.

As a specific example, in the following we consider an object comprising $10^6$ atoms. Let us first emphasize that even prior to any probing of gravity, a successful SGI will already achieve 3 orders of magnitude more atoms than the state of the art in macroscopic-object interferometry \cite{Arndt}, thus contributing novel insight to the foundations of quantum mechanics. Another contribution to the latter would be the ability to test continuous spontaneous localization (CSL) models (e.g. \cite{Romero8} and references therein).

When considering gravity, the first contribution of such a massive-object SGI would simply be to measure small $g$ (the local gravitational acceleration). Such an experiment would first of all enable to verify the created superposition of the macroscopic object \cite{Scala2013,BoseNanoObject,Marco-Barker2020}.
A second contribution of such a SGI in the field of gravity would be in testing modifications to gravity at short ranges (also known as the fifth force), as one of the SGI paths may be brought close to a massive object, thus allowing to probe gravity at short distances\,\cite{short-range-gravity}. This is a regime in which light-based interferometers would be difficult to use due to diffraction and light scattering near the surface.
Once SGI technology allows for large masses, a third contribution will be the testing of hypotheses concerning gravity self-interaction \cite{self-gravity, self-gravity2, self-gravity3}, and when large-area SGI with large masses is available, a fourth contribution would be to detect gravitational waves \cite{MIMAC}. Finally, it is claimed that placing two such SGIs in parallel next to each other will allow probing the quantum nature of gravity \cite{Bose2017,Marletto2017}. Let us emphasize, that although high accelerations may be obtained with multiple spins, in the following we discuss only the case of a macroscopic object with a single spin, as the observable of such a quantum-gravity experiment is entanglement between two spins, and averaging over many spins may wash out the signal. Furthermore, a multi-spin SGI would give rise to multiple trajectories. 

We focus solely on the required SG parameters, which include the necessary accelerations and accuracy in relation to the required splitting and coherence length, respectively. Other issues such as material science (e.g. clean surfaces against patch potentials, spin contaminations, or spin coherence time near the surface), or the very good vacuum which will be required, will not be dealt with here. We will also not deal with the issue of the Casimir-Polder or diamagnetic interactions that may arise. These issues have been dealt with extensively in previous works \cite{Casimir,Bose2017}. While we believe all discussed applications noted here, including the two parallel interferometers for the quantum-gravity experiment, may be done in the longitudinal configuration presented in the experimental part of this paper, transverse (i.e. 2D) interferometers may also be realized with the same techniques discussed here. Both the 1D and 2D interferometers involve similar operations and present the same challenges, and in the following we will not discuss the differences between them.

We consider, for example, a nano-diamond composed of $10^6$ carbon atoms (corresponding to a sphere radius of 11\,nm) with a single  NV spin embedded in it. Coherence time of 1\,ms has been demonstrated at room temperature\,\cite{NVcoherence1}, and 0.6\,s while cooling to 77\,K\,\cite{NVcoherence2}. The total interferometer time of the proposed experiment will have to be shorter than these times. Utilizing well-known NV techniques (e.g. see for example our own work\,\cite{NV1,NV2}), we do not see any fundamental spin-related obstacles.

The experimental procedure is the same as for the atomic full-loop SGI demonstrated above, with the required adjustments for manipulating the nano-diamond. The experiment begins by trapping and cooling the center-of-mass motion of the nano-diamond\,\cite{CoolingGroundState, CoolingGroundState2}. We note that to the best of our knowledge, ground-state cooling of nano-diamonds is yet to be achieved\,\cite{NVMagneticTrap}. We then release the nano-diamond from the trap, and prepare it in a spin superposition of the $\pm1$ spin projections using a $\pi/2$ microwave pulse (a two-photon transition). Placing the nano-diamond 1\,$\mu$m away from a $1\times1\,\mu$m wire on an atom chip, carrying 1\,A of current (with a current density of $10^8$\,A/cm$^2$, achievable with CNT embedded Cu wires \cite{CNT}), we get a magnetic gradient of $8.7\times10^4$\,T/m. For $10^6$ carbon atoms, the acceleration for a single spin (1 Bohr magneton) is $a=\mu_B \partial_z B/m = 81\,$m/s$^2$. We then apply a microwave $\pi$ pulse to inverse the relative acceleration, and apply the stopping and reversing pulses $T_2$ and $T_3$. After another $\pi$ pulse, we apply $T_4$, and stop the relative motion (see Fig.\,1 for description of pulses). For total interferometer times of $T = (0.1,\,1,\,10,\,100)$\,ms, we get maximum splittings of $\Delta z = a(T/4)^2 = 5\times(10^{-8},\,10^{-6},\, 10^{-4},\, 10^{-2})$\,m. Maintaining a constant acceleration for the long durations requires placing current-carrying wires along the wavepackets' trajectories such that they are always in a region of strong gradient (e.g. as previously suggested in Ref.\,\cite{MIMAC}), or realizing the experiment in a very strong quadrupole field created by coils (e.g. superconducting). We note that if the nano-diamond is very close to the wire, one of the spin states will have to be $0$ so that it does not crash into the wire.
The spin state is then read out by standard NV optical techniques. Let us briefly recall that the phase is predicted to be independent of the initial motional condition\,\cite{BoseNanoObject}.

The crucial parts of the experiment are the ground-state cooling and the recombination.
The required recombination accuracy is on the order of the coherence length, given by the harmonic oscillator length $\sqrt{\hbar/m\omega}$, which for the considered object is assumed to be about 0.1\,nm (assuming ground-state cooling with $\omega /2\pi = 80\,$kHz, as in Ref.\,\cite{CoolingGroundState}). This value can be increased by adiabatically lowering the trap frequencies after cooling. We do not take into account some works which claim that techniques exist with which the coherence length may be increased even further \cite{ketterle,Romero6}, as these are done for spatial interference fringes, where the ``local" coherence length is what matters\,\cite{ketterle}. In contrast, what matters for the full-loop SGI is the overlap integral, for which the coherence length only depends on the initial momentum width, namely on the ground state energy when cooling to the ground state.
This recombination accuracy represents the main technical challenge, as it must be better than the achieved coherence length. In the experimental results presented above for an atomic SGI, high visibility is achieved for a 700\,nm coherence length, which allows us to assume we have obtained a recombination accuracy on the order of 100\,nm (when the maximal splitting was an order of magnitude larger). Improving this recombination accuracy by 3 orders of magnitude (while maintaining the same ratio to the maximal splitting) is well within reach by utilizing better current sources with less current and time jitters. (We note that at present we use current sources with instabilities far above shot-noise.)

Let us briefly also touch upon the topic of decoherence due to external and internal degrees of freedom, namely, which-path information due to some scattering event between the environment and the nano-diamond. The information may be encoded in the environment (i.e. the scattered particle, equivalent to the interaction with some noise with a correlation length smaller than the spatial splitting \cite{imry1990}), or in the internal degrees of freedom.
For example, as explained in Ref. \cite{Decoherence}, any object with excited internal degrees of freedom may emit radiation which would localize it. Other sources of decoherence may originate from any differential interaction the nano-diamond suffers between the two paths.

As a simple example of lack of decoherence due to the internal degrees of freedom, we can consider the atom interferometer demonstrated in this paper.
The atom has many internal degrees of freedom such as those of the electrons or the nucleus. Decoherence can occur due to the emergence of which-path information, or in other words orthogonality. However, if the inner degrees of freedom do not develop
some orthogonality along the two paths, then they are irrelevant for the interferometer, as no which-path information is encoded in them.
What can be relevant is some differential interaction between those internal degrees of freedom and the environment which can create such orthogonality, where by differential we mean that the interaction with the environment affecting the internal degrees of freedom, in a coordinate system relative to the c.o.m, is different for the two paths of the interferometer. In the case of atom interferometry, it seems that external perturbations are negligible with respect to the atom's bare Hamiltonian.

In contrast, with the nano-diamond the differential interaction between the internal degrees of freedom and the environment may not be negligible. For example, a background gas collision can introduce rotation or phonons in the nano-diamond, in only one of the paths. The collision can even be with a cosmic muon.
Furthermore, an external magnetic/electric/EM field that creates rotation or phonons, or some spin flip of a contamination spin, just in one of the paths (due to a small correlation length), would give rise to orthogonality. A more subtle process of decoherence would be the excitation of phonons, solely due to the magnetic force acting on the NV centers. Here the sudden (non-adiabatic) force acting on the spin would create a phonon in the lattice (this may be estimated through a Debye-Waller factor). As the acceleration in the two paths is different, the excited phonons would be different, giving rise to orthogonality. Even if the acceleration is symmetric but opposite in sign, the opposite $k$-vector of the identical phonons would create orthogonality. To maintain adiabaticity, one may have to resort to engineering a gradual increase of the magnetic gradient within the pulse duration.

Calculating the cross-section/probability for such events and the amount of orthogonality they create is a daunting task. Some of these calculations have already been done in Refs. \cite{Bose2017,Romero6} and references therein.
In any case, we do not see a clear correlation between these cross sections and the internal temperatures, and so a priori there is no clear need for the cooling of the internal degrees of freedom or the environment, as long as the wavelength of the blackbody radiation is larger than the splitting induced in the SGI.
If, however, for some reason, cooling of the internal degrees of freedom is required, the experiment may be done in a dilution fridge cryostat, with which the atom chip technology is compatible. Additional methods to avoid decoherence, such as dynamic decoupling, have been suggested as well \cite{Dynamic}. In any case, the fact that the SGI can achieve large spatial splitting in a very short time, as presented above, is a crucial advantage when addressing the issue of decoherence. 

Finally, let us emphasize that the SGI also has several clear advantages over laser-pulse interferometers in the context of such an experiment:
(a) Laser-pulse matter-wave beam splitters require an appropriate internal transition to coherently scatter photons; this demand severely restricts the applicability of these splitters to macroscopic objects. The SGI merely requires a magnetic moment, which enables to readily achieve the magneto-mechanical coupling.
(b) No heating is generated on the macroscopic object, since no light is scattered.
The scattering of light may also severely reduce the amount of light interacting with the two-level system embedded in the object. Lack of light scattering also suppresses the decoherence rate.
(c) In all experiments not dealing with the entanglement, as the number of spins can be increased with the macroscopic object's mass (typical concentrations are in the parts per million), the acceleration is independent of the mass, enabling to split large masses, or to achieve large angle interferometry for small masses.

\section*{Conclusion}

To conclude, we have demonstrated for the first time a full-loop SGI, consisting of freely propagating atoms exposed to magnetic gradients, as originally envisioned decades ago. We have unambiguously shown recombination in both momentum and position. We have shown that SG splitting may be realized in a highly coherent manner with macroscopic magnets without requiring cryogenic temperatures or magnetic shielding. Furthermore, we have analyzed the limits of our system's accuracy.

We briefly compare our experiments to state-of-the-art SG interferometry\,\cite{OldSG0,OldSG2,OldSG3,OldSG4,OldSG5,OldSG6,OldSG7,OldSG8, Marechal2000,OldSG9,OldSG10}.
While these longitudinal beam experiments did show spin-population interference fringes, the experiments presented here are very different. Most importantly, as explained in\,\cite{OldSG6} and \cite{Marechal2000}, the full-loop configuration was never realized, as only splitting and stopping operations were applied (i.e., no active recombination); namely, wavepackets exit the interferometer with the same separation as the maximal separation achieved within.

Finally, achieving this high level of control over magnetic gradients may facilitate fundamental research as well as technological applications. Specifically, we show that in principle, full-loop SG interferometry with a macroscopic body is feasible. It may be used to test the foundations
of quantum theory, as well as to probe exotic forces such as the fifth force. It may be developed as a gravitational sensor, serve to test exotic gravitational models such as self-interaction, and finally, may probe the quantum nature of gravity.

\section*{Methods}

\subsection{Detailed experimental scheme}
In the following we describe our experimental sequence. We begin by preparing a BEC of about $10^4$ $^{87}$Rb atoms in the state $\vert F, m_F\rangle =\vert 2,2\rangle$ in a magnetic trap located around $91-96\,\mu$m $\pm 1\,\mu$m below the atom chip surface (different experiments use different initial positions). The harmonic frequencies of the trap are $\omega_x/2\pi = 38$Hz and $\omega_y/2\pi\approx\omega_z/2\pi = 127$Hz. The trap is created by a copper structure located behind the chip with the help of additional homogeneous bias magnetic fields in the $x$, $y$ and $z$ directions. The BEC is then released from the trap, and falls a few $\mu$m under gravity for a duration $T_{d0}= 0.9-1.4\,ms$ (see Fig.\,\ref{fig:full loop timing} for a timing diagram). During this time the magnetic fields used to generate the trap are turned off completely. Only a homogeneous magnetic bias field of 36.7\,G in the $y$ direction is kept on to create an effective two-level system via the non-linear Zeeman effect such that the energy splitting between our two levels $\vert 2,2\rangle\equiv|2\rangle$ and $\vert 2,1\rangle\equiv|1\rangle$ is $E_{21}$ $\approx h\times$25\,MHz, and where the undesired transition is off-resonance by $E_{21}-E_{10}  \approx h\times $180\,kHz. As the BEC expands, interaction becomes negligible, and the experiment may be described by single-atom physics.

Next, we apply a radio-frequency (RF) $\pi/2$ pulse (10\,$\mu$s duration) to create an equal superposition of the two spin states,  $|1\rangle$ and $|2\rangle$, and a magnetic gradient pulse (splitting pulse) of duration $T_1=4-40\,\mu$s which creates a different magnetic potential for the different spin states $m_F$, thus splitting the atomic wave packet into two wave packets with different momenta. The chip wire current is driven using simple 12\,V batteries connected in series, and is modulated using a home-made current shutter. The SG acceleration is in the range of 59.5 - 2641 m/s$^{2}$ (depending on chip current). The acceleration is measured by splitting the wavepakcets using a single pulse, and measuring the relative distance as a function of the time-of-flight\,\cite{machluf}.

After a delay time $T_{d1}$, we apply a stopping pulse of duration $T_2$ that cancels the relative momentum of the two wavepackets, and immediately after a gradient pulse for accelerating the atoms back towards each other ($T_3$). After an additional delay time $T_{d2}$ we apply a final stopping pulse ($T_4$) so that the two wave packets overlap in momentum and position. %These four gradient pulses occur in between the two RF $\pi/2$ pulses.

A second RF $\pi/2$ pulse is applied only at the time of measurement (meaning the two wave packets have a different spin throughout the propagation), which completes the interferometric sequence. Without the magnetic gradient pulses and their effect on the spatial wavefunction, the two $\pi/2$ pulses correspond to a Ramsey sequence. Our signal is an interference pattern formed by measuring the spin population (e.g. starting with $S_{\tilde{x}} = +1$ and measuring $S_{\tilde{z}}$).

The visibility, or contrast, which represents spin coherence in our experiment, is measured by changing the phase between the two interferometer arms, and observing the resulting population oscillation between the two states. In some measurements, we scan the magnetic gradient pulse duration, thus adding a relative phase between the arms\,\cite{T3SGI} (as in Figs.\,\ref{fig:optimization}, \ref{fig:recombination}).

In other measurements in which the gradient pulse durations are fixed, we create a population interference pattern by shifting the phase $\phi$ of the last RF $\pi/2$ relative to the first RF $\pi/2$ pulse. This creates population oscillations between the states (as a function of $\phi$), generating the oscillations shown in the insets of Fig.\,\ref{fig:contrast}. Contrast is then evaluated by fitting the population fringes to a function of the form $P(\phi) = 0.5 C \sin(\phi+\phi_0)+const$, where $C$ is the contrast, $\phi$ is the applied phase shift, and $\phi_0$ is a constant phase term. As noted, the resulting contrast shown in Fig.\,\ref{fig:contrast} is normalized to that of a pure Ramsey sequence, i.e. a sequence without any magnetic gradients. This cancels technical effects that reduces the contrast, such as non-zero RF detuning (causing imperfect $\pi/2$ pulses), spin decoherence due to external magnetic noise, etc..

As noted before, we add one or two RF $\pi$ pulses in between the two $\pi/2$ pulses, giving rise to an echo sequence which suppresses the dephasing taking place due to magnetic noise and inhomogeneous magnetic fields in our chamber. This allows us to increase the spin coherence time from $\sim 400$ $\mu$s up to $\sim 4$ ms.

Relative populations in each output port are measured by applying a homogeneous magnetic gradient that separates the spin states, and counting the number of atoms in each output state using standard absorption imaging. The homogeneous gradient is created by running a current in the copper structure behind the chip for a few ms.

Finally, it is worth noting that while all gradient pulses come from the same chip wires, the magnetic pulses may be considered as an analogy of the original thought experiment in which there were different spatial regions with different permanent magnets. This is so as in each pulse the current and duration may be different and have individual jitter, and in addition the atom position and consequently the gradient are different.

\subsection{Full-loop configurations}
\begin{figure}
\centerline{
\includegraphics[width=\columnwidth]{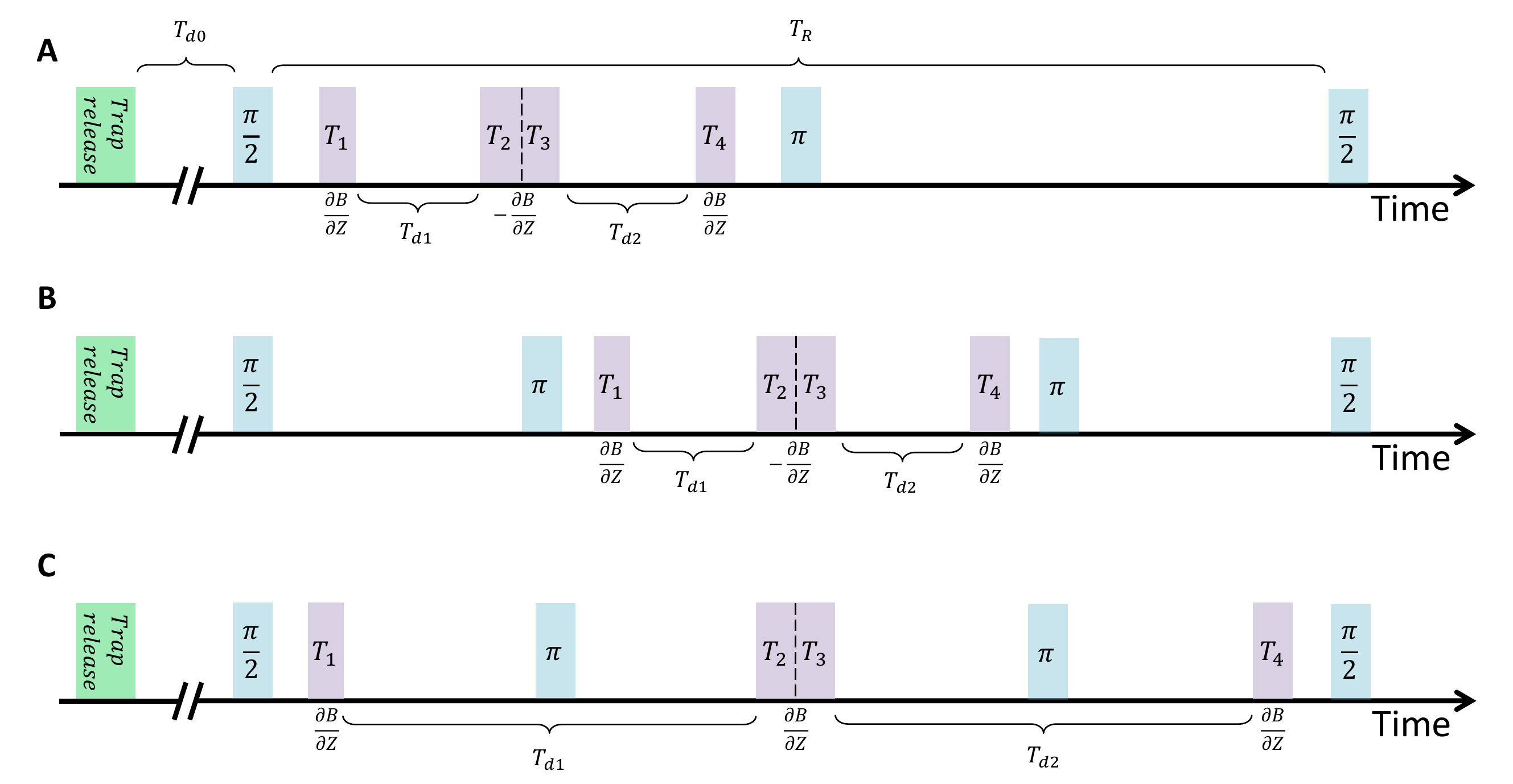}}
\caption{\label{fig:full loop timing} Timing diagram of the full-loop experimental sequences. (A),(B) The current inversion scheme, where the sign of the gradient $\partial B/ \partial z$ is switched during the sequence. The difference between (A) and (B) is the additional $\pi$ pulse before the gradients, used to increase the coherence time. (C) The spin inversion scheme, in which the sign of the gradient $\partial B/ \partial z$ is kept constant, and the relative force between the spin states is inverted by using the $\pi$ pulses in between the gradients, inverting the spin states.}
\end{figure}

Here we describe in more detail the 'current inversion' and 'spin inversion' sequences shown in a timing diagram in Fig.~\ref{fig:full loop timing}, which are used to generate the data in Fig.\,\ref{fig:contrast}.
In the first sequence, after applying a $\pi/2$ pulse and the splitting gradient $T_1$ in one direction (downwards towards gravity), we reverse the sign of the acceleration by reversing the sign of the currents in the chip wires for the stopping and reversing gradients $T_2$ and $T_3$, working in the opposite direction (upwards towards the chip, this is done using two independent current shutters connected to the chip in opposite directions). The opposite gradient causes the relative movement between the wave packets to stop during $T_2$, and to change sign during $T_3$. Finally the wave packets are brought to a relative stop and spatial overlap by the second stopping pulse $T_4$ given in the same direction as the first gradient. The sequence is finished by applying a $\pi$ pulse (to increase coherence time, giving rise to an echo sequence) and a $\pi/2$ pulse, for mixing the different spin states and enabling spin populations interference (the $\pi/2-\pi-\pi/2$ sequence is symmetric in time). The four consecutive gradient pulses used in the current inversion sequence are applied either after the first $\pi/2$ pulse (i.e. only a single $\pi$ pulse is used as described above), or in between two $\pi$ pulses to further increase the coherence time (the $\pi/2-\pi-\pi-\pi/2$ sequence is again time symmetric).

In the second sequence, we keep the same current direction in all gradient pulses, while reversing the spins with the help of two $\pi$ pulses. These pulses are applied, first just before the stopping gradient pulse $T_2$ and second just after the reversing gradient pulse $T_3$. Each $\pi$ pulse causes the spin states to flip sign, thus changing the direction of the applied momentum kicks in the center-of-mass frame (in lab frame, all gradient pulses push the atoms downwards towards gravity). Both the current inversion and spin inversion sequences are used to generate the data points for Fig.\,\ref{fig:contrast}.

\subsection{Minimizing the visibility loss due to momentum splitting (Fig.\,\ref{fig:recombination})}

In Fig.\,\ref{fig:recombination}, we validate that the loss of visibility is due to spatial splitting, by minimizing the visibility loss due to momentum splitting for each value of $T_{d1}$, using the following optimization procedure. We scan for the optimal value of $T_2$, i.e. the value which obtains the maximal visibility, for several delay times ($T_{d1}=100,\,200,\,300,\,400\,\mu$s), in a similar way to what is shown in Fig.\,\ref{fig:optimization}. We then determine the optimal $T_2$ for any given value of $T_{d1}$ using a polynomial interpolation. The blue data in Fig.\,\ref{fig:recombination} are taken by using the optimal values of $T_2$ as a function of $T_{d1}$. This ensures that the visibility loss due to momentum splitting is minimized for every value of $T_{d1}$, meaning we see visibility loss mostly due to spatial splitting.

\subsection{Calculation of $l_z$}
In the fit of the blue data of Fig.\,\ref{fig:recombination}, $T_d = \tau$ represents the time at which the contrast drops to $1/\sqrt{e}$ of its full value. Accordingly,
we estimate the spatial coherence length $l_z$ from the spatial splitting at $T_d=\tau$.
This is the sum of three contributions: 1. the splitting under constant acceleration $a$ for a duration $T_1$, given by $\frac12 a T_1^2$; 2. the splitting at constant velocity $a T_1$ for a duration $T_d$, given by $a T_1 T_d$; 3. the splitting under constant deceleration $-a$, from initial velocity $a T_1$ until zero relative velocity, for duration $T_2=T_1$, given by $a T_1 T_1 -\frac12 a T_1^2 = \frac12 a T_1^2$. Summing all contributions and setting $T_d=\tau$, we obtain $l_z = aT_1^2+aT_1\tau$. Using the experimental parameters $a=481.6$\,m/s$^{2}$, $T_1=5.4\,\mu$s, and the fit result of $\tau = 186.8\pm22.5\,\mu$s, we obtain $l_z = 0.5 \pm 0.07\,\mu$m.

\subsection{Coherence scales}

Here we compare the experimental results of the coherence scales to those obtained from Eq.\,\ref{eq:coherence length}. Assuming the BEC is a minimal-uncertainty wavepacket, at trap release time  both its coherence scales are set by a single number - the size of the BEC wavepacket. We therefore begin by calculating this number and validating it experimentally. After some time-of-flight, however, the BEC momentum width increases due to conversion of the mean-field potential energy to kinetic energy\,\cite{Phillips}, which we account for using the theory of Ref.\,\cite{Japha2019}.  %We then compare the results of Eq.\,\ref{eq:coherence length} to the experimental values.

The in-trap Thomas-Fermi condensate half-length $w_0$ in the z (gravity) direction is given by \cite{Ketterle1999} $w_0 = \sqrt{2\mu/m}/\omega_z$, where the Thomas-Fermi expression for the chemical potential $\mu$ for an harmonically confined condensate is given by \cite{Ketterle1999}
$\mu^{5/2} = 15\hbar^2m^{1/2} N \bar \omega^3 a / 2^{5/2}$,
and where $\bar \omega = (\omega_x \omega_y \omega_z)^{1/3}$ is the geometric mean of the trap frequencies, $N$ is the number of atoms, and $a=5.18$\,nm is the $^{87}$Rb s-wave scattering length. For our experimental parameters of $N=10,000$ atoms, $\omega_{x,y}/2\pi=40$\,Hz and $\omega_z/2\pi=126$\,Hz, we obtain $w_0 = 2.88\,\mu$m.

We also experimentally validate this value: as imaging the wave packet at short time-of-flight $T_{d0}$ [see Fig.\,\ref{fig:full loop timing}] does not give reliable results due to refraction effects around a BEC with high optical density, we measure the Thomas-Fermi wave packet size as a function of $T_{d0}$, and use the known trap frequency $\omega_z$ to calculate $w_0$ according to $w(t) = w_0\sqrt{1+\omega^2 T_{d0}^2}$ \cite{Ketterle1999}. This method gives the value $w_0 = 3.04 \pm 0.3\,\mu$m, in reasonable agreement with the theoretically calculated value.

As the HD theory and the generalized wave-packet model\,\cite{Japha2019} both assume a Gaussian wavepacket, we express the Thomas-Fermi size in terms of Gaussian width by fitting it to a Gaussian profile, which gives $\sigma_z \simeq 0.41 w_0 = 1.2\,\mu$m for the in-trap size. Also taking into account the experimental time-of-flight of $T_{d0}=$1\,ms (i.e. the time difference between trap release and the start of the SGI sequence), the wavepacket size at the start of the SGI sequence is $\sigma_z(t=0)=\sqrt{1+\omega_z^2 T_{d0}^2}=1.5\,\mu$m.

The measured values of $l_p$ and $l_z$ can now be compared to theory (Eq.\,\ref{eq:coherence length}) using the obtained BEC wavepacket size. A momentum coherence width of $l_p/m=0.12\,$mm/s corresponds to a wavepacket size of $\sigma_z = 6.1\,\mu$m. This means that the momentum coherence width is about 4.1 times narrower than expected for a pure condensate. The discrepancy could originate from the contribution of the thermal fraction of the atoms, although we evaluate the BEC fraction to be larger than 0.7, so we cannot fully explain this effect yet.

A coherence length of $l_z=0.5\,\mu$m corresponds to a momentum uncertainty of $\sigma_p/m= 2$\,mm/s, while the expected momentum uncertainty of a pure BEC after 1\,ms expansion is $\sigma_p/m\approx 0.6$\, mm/s (including increased momentum width due to conversion of the mean-field potential energy to kinetic energy). This means that the coherence length is shorter than expected for a pure condensate, a discrepancy which is only partially explained by the thermal fraction of atoms.

%\\
%\\

\begin{acknowledgments}
We thank Mark Keil for helpful discussions. We are grateful to Zina Binstock for the electronics, and the BGU nano-fabrication facility for providing the high-quality chip. This work is funded in part by the Israeli Science Foundation (grants No. 1314/19 and 1381/13), the EC ``MatterWave” consortium, the DFG through the DIP program (FO 703/2-1), and the program for postdoctoral researchers of the Israeli Council for Higher Education. A.M. is supported by Netherlands Organization for Scientific Research (NWO) Grant number 680-91-119. SB would like to acknowledge UK EPSRC grants EP/N031105/1 and  EP/S000267/1.
\end{acknowledgments}

%\bibliography{SGIRefs}

\end{document}